\documentstyle [11pt,aaspp4,epsf]{article}
\newcommand{\mincir}{\raise
-2.truept\hbox{\rlap{\hbox{$\sim$}}\raise5.truept\hbox{$<$}\ }}
\newcommand{\magcir}{\raise
-2.truept\hbox{\rlap{\hbox{$\sim$}}\raise5.truept\hbox{$>$}\ }}
\newcommand{\minmag}{\raise
-2.truept\hbox{\rlap{\hbox{$<$}}\raise6.truept\hbox{$<$}\ }}
\newcommand{\be}{\begin{equation}}
\newcommand{\ee}{\end{equation}}
\newcommand{\ba}{\begin{eqnarray}}
\newcommand{\ea}{\end{eqnarray}}
\newcommand{\brr}{\begin{array}}
\newcommand{\err}{\end{array}}
\newcommand{\bc}{\begin{center}}
\newcommand{\ec}{\end{center}}

\begin{document}

\title{The X--ray Cluster Dipole}
\author{M. Plionis$^{1}$ \& V. Kolokotronis$^{2}$}
\affil{$^1$ National Observatory of Athens, Lofos Nimfon, Thesio, 18110 Athens,
Greece}
\affil{$^2$ Astronomy Unit, School of Mathematical Sciences,
Queen Mary and Westfield College, Mile End Road, London E1 4NS}

\received{---------------}
\accepted{---------------}


\begin{abstract}
We estimate the dipole of the whole sky X--ray flux--limited sample of 
Abell/ACO clusters (XBACs) and compare it to the optical cluster dipole which 
is known to be well aligned with the CMB dipole and which converge to its 
final value at $\sim 160h^{-1}\,$Mpc (Branchini \& Plionis 1996 and 
references therein). The X--ray cluster dipole is well aligned ($\,\mincir 
25^{\circ}$) 
with the CMB dipole, while it follows closely the radial profile of its 
optical cluster counterpart although its amplitude is $\sim 10 - 30$ per 
cent lower. 
In view of the fact that the the XBACs sample is not affected by the 
volume incompleteness and the projection effects 
that are known to exist at some level in the optical parent Abell/ACO 
cluster catalogue, our present results confirm the previous optical 
cluster dipole analysis that there are significant contributions to the
Local Group motion from large distances ($\sim 160h^{-1}\,$Mpc).
In order to assess the expected contribution to the X--ray cluster dipole from
a purely X--ray selected sample we compare the dipoles of the XBACs and the  
Brightest Cluster Sample (Ebeling et al. 1997a) in their overlap region. 
The resulting dipoles are in mutual good aggreement
with an indication that the XBACs sample slightly underestimates the
full X--ray dipole (by $\,\mincir$ 5 per cent) while the Virgo cluster
contributes about 10 - 15 per cent to the overall X--ray cluster dipole.
Using linear perturbation theory to relate the X--ray cluster dipole to the 
Local group peculiar velocity we estimate  $\beta_{\rm c_{x}}
(\equiv \Omega_{\circ}^{0.6}/b_{\rm c_{x}}) \simeq 0.24 \pm 0.05$.
\end{abstract}


\section{Introduction}
A lively debate has been going on the recent years on which is the spatial 
extent of the distribution of mass inhomegeneities that cause the LG motion. 
Assuming gravitationally instability as the cause of cosmic motions and
using as tracers of the matter distribution optical and IR galaxies,
many studies (cf. Yahil, Walker \& Rowan-Robinson 1986; Harmon et
al. 1987; Lahav 1987; Lahav, Rowan-Robinson \& Lynden-Bell 1988;
Lynden-Bell, Lahav \& Burstein 1989; Strauss et al. 1992; Hudson 1993)
have shown that most, if not all, of the peculiar acceleration of the
LG is induced within 40 - 50$h^{-1}\,$Mpc. 
Other analysis of galaxy samples have presented
indications, of varying strength, for contributions from much larger
depths, ranging from $\sim$ 100$h^{-1}\,$Mpc to $\sim$
150$h^{-1}\,$Mpc (Plionis 1988; Rowan-Robinson et al. 1990; Plionis,
Coles \& Catelan 1993; Vasilakos \& Plionis 1997). However the
difficulty with such studies in providing a definite answer is that
the galaxy samples are not volume limited but rather magnitude- or
flux--limited which introduces an inherent uncertainty due to the rapid
decrease of their selection function with distance from the observer.  

Alternatively, galaxy clusters being the largest gravitationally-collapsed 
structures in the universe and luminous enough to be volume-limited out to 
large distances have also been used to probe the local acceleration field.
Existing studies, all based on the optically selected Abell/ACO
clusters (Abell, Corwin \& Olowin 1989) provide strong
evidence that the LG dipole has significant contributions from depths
up to $\sim 160h^{-1}\,$Mpc (Scaramella, Vettolani \& Zamorani 1991; 
Plionis \& Valdarnini 1991; Branchini \& Plionis 1996). 
However, due to the the volume incompleteness of richness class R=0 clusters 
(cf. Peacock \& West 1992) and to optical projection effects (enhancement of 
galaxy density along the direction of foreground rich clusters which cause 
inherently poor background clusters or groups to appear rich enough to be
included in the sample), these results should be verified by well
defined cluster samples, free of such biases.    

In the X--ray band the physical reality of clusters is unquestionable 
due to their strong ICM X--ray emission. Two large X--ray cluster
samples have been recently compiled; the XBACs sample by Ebeling et al. (1996),
from carefully cross-correlating the {\small ROSAT} all-sky X--ray survey
(Tr$\ddot{\rm u}$mper 1990; Voges 1992) with the Abell/ACO cluster sample
and the Brightest Cluster Sample (BCS) by Ebeling et al. (1997a) from an 
additional cross-correlation of the RASS sources with the Zwicky cluster
catalogue but it also contains clusters purely selected in X-rays (for the 
complete definition see section 2.2). The XBACs sample provides, for the 
first time, a whole sky, flux--limited, sample of X-ray galaxy clusters 
suitable for investigating the local
acceleration field (for an early attempt using mostly HEAO-1 data see Lahav
et al. 1989). The BCS sample covers only the northern sky and it 
has been used to investigate the evolution of the X--ray cluster luminosity 
function (Ebeling et al. 1997b) while both samples will be useful for 
establishing, among other things, the cluster correlation function 
(Edge et al. in preparation). Nevertheless, both catalogues suffer
from some degree of incompleteness at low galactic latitude (see also
sections 2.3 and 3 for relative corrections).
Apart from these two samples, other X--ray cluster samples, also 
based on {\small ROSAT} data, are under compilation, most notably the
ESO KP catalogue for southern clusters (Collins et al. 1995; de Grandi
1996; Guzzo et al. 1995, 1996).

As in the case of all flux--limited samples, the use of these X--ray
cluster samples to investigate the very distant contributions to the local
acceleration field is limited exactly due to their flux--limited nature.
In fact Kolokotronis et al. (1997) found, using numerical 
experiments, that such samples will underestimated the true underlying cluster
dipole by $\sim$ 15 per cent on average if such distant contributions do 
exist. We further caution the reader that using clusters to estimate 
the local acceleration field maybe problematic because:
\begin{itemize}
\item Clusters may not trace well the very local gravity field due to their 
large intercluster separation ($\sim 30 - 40h^{-1}\,$Mpc) and unless the
local velocity field is cold, which does seem to be the case (cf. Peebles 
1988), attempts to relate the cluster dipole with the LG peculiar velocity 
could give erroneous results.
\item Existing cluster samples are incomplete in many different ways. 
For example the Virgo cluster is missing from the optical Abell/ACO catalogue
and thus also from the XBACs sample. Furthermore, the present X--ray 
cluster samples may suffer from incompleteness in the nearby Universe due
to problems in reliably detecting extended low-surface brightness emission.
\end{itemize}
These limitations will be investigated by comparing, in their overlap region,
the XBACs and BCS dipoles, since the latter sample is nearer to being purely 
X--ray selected and it also contains the Virgo cluster.

The outline of this paper is as follows. The X--ray samples and various 
selection biases are discussed in section 2. The main dipole results 
are presented in section 3, while in section 4 we estimate
the cosmological $\beta$ parameter. Finally, our main conclusions are 
presented in section 5.

\section{X--ray samples \& selection effects}
Both X--ray samples consist of clusters identified in the {\small ROSAT} all 
sky survey (RASS) by a combination of two detection algorithms (the
Standard Analysis Software System and Voronoi Tesselation Percolation;
SASS and VTP hereafter) for fluxes above a particular flux limit 
($S_{\rm lim}$).
The use of the VTP identification algorithm allows quite reliable cluster 
detections even at low redshifts and
improves greatly the flux determination for the X--ray sources,
initially misassessed by SASS (see Ebeling et al. 1996; 1997a and
references therein for superiority of VTP over SASS technique). 

\noindent
Throughout this work we will be using the following definition of 
distance (Mattig's formula): 
$$ r=\frac{c}{H_{\circ} q_{\circ}^{2} (1+z)} \left[ q_{\circ} z + (1-q_{\circ})
(1-\sqrt{2 q_{\circ} z +1}) \right] $$
with $q_{\circ}=0.5$ and $H_{\circ}=100h$ km s$^{-1}\,$Mpc$^{-1}$.

\subsection{XBACs sample}
The XBACs sample consists of the X--ray brightest Abell/ACO clusters 
that have been detected in the {\small ROSAT} all sky survey (RASS)
for fluxes above $S_{\rm lim}=5 \times 10^{-12}\,$erg
s$^{-1}\,$cm$^{-2}$ (0.1 - 2.4 keV) with redshifts limited by $z\le 
0.2$. The sample contains 253 clusters out of which 242 have
$|b|\ge 20^{\circ}$ and thus
it is the largest X--ray flux--limited cluster sample to date (though not 
entirely X--ray selected). 
The X--ray fluxes measured initially by the SASS point source
detection algorithm are superseded by VTP measurements that
account for the extended nature of the emission. In addition, the
difficulty of the SASS algorithm to actually detect nearby X--ray
emission has been mostly corrected by running VTP on the {\small RASS}
fields centered on the optical positions of all nearby Abell/ACO
clusters ($z\le\,$0.05) irrespective of whether or not they are
detected by SASS. Ebeling et al. (1996) estimate, after a careful
analysis of possible selection effects and biases that the overall
completion rate of this X--ray sample is more than 80 per cent.   

\subsubsection{XBACs systematic effects}

Due to the cross-correlation of the RASS with the Abell/ACO
cluster positions, it is very probable that the systematic biases 
from which the latter suffer, could also creep in the XBACs
sample. Ebeling et al. (1996) have shown that the XBACs flux--limited
sample is free of the known volume incompleteness, as a function of
distance, of the richness class R=0 Abell/ACO clusters, exactly
because of the flux--limited property of the XBACs sample which is
such, that it contains at large distances the inherently brighter and
thus richer Abell/ACO clusters for which there is no volume
incompleteness.  

Another bias from which the optical clusters suffers and which could
therefore affect also the XBACs sample, is the significant
distance dependent density variations between the northern Abell 
and southern ACO parts of the combined cluster sample 
(cf. Batuski et al. 1989; Scaramella et al. 1990; Plionis \& Valdarnini 1991).
These density variations are most probably due to the higher sensitivity of 
the ACO IIIa-J emulsion plates which results in detections of inherently 
poorer nearby ACO clusters. As a first step to quantify the overall magnitude 
of the effect on the XBACs sample we estimate, for $|b|>30^{\circ}$, the 
density ratio and its Poisson error between the Abell and ACO parts of the 
sample within the volume limited region of the optical cluster sample;
\[
\frac{\bar{n}_{\rm ACO}}{\bar{n}_{\rm Abell}}(r < 240 {\it h}^{-1}\, 
\mbox{ Mpc})  
\simeq \left\{ \begin{array}{ll}
 	1.56 \pm 0.14 & \mbox{ optical} \\
	1.12 \pm 0.17 & \mbox{ XBACs}
	 \end{array} \right. 
\]
where each cluster has been weighted by ${\cal P}(b) (\equiv
10^{{\cal A} cosec|b|})$,
to account for the number density decrease due to Galactic absorption
\footnote{The amplitude of this function has been estimated from each 
individual cluster sample and it is consistent with the usually quoted values 
(${\cal A}_{\rm Abell}\simeq -0.3$ and ${\cal A}_{\rm ACO}\simeq
-0.2$).}. It is evident that the X--ray selection has corrected the
significant systematic density variation seen in the optical
sample. The lower {\small X--ray} detection rate of ACO clusters is
probably because they are inherently poorer clusters (and thus weak
X--ray emitters), revealed due to the higher sensitivity of the IIIa-J
emulsion ACO plates. 

As already mentioned, the apparent density variations between the Abell 
and ACO samples are distance dependent (Plionis \& Valdarnini 1991) 
and since uncorrected systematic density differences between two parts 
of the sky can introduce spurious contributions to the dipole, we will 
statistically correct such variations by weighting each Abell 
cluster with:
\begin{equation}\label{eq:weight}
w(r,{\rm \delta} V_{\rm i})=\frac{\bar{n}_{\rm ACO}(r,{\rm \delta}
V_{\rm i})}{\bar{n}_{\rm Abell}(r,{\rm \delta} V_{\rm i})}
\end{equation}
The robusteness of our results will be  checked by using a large 
number of bin sizes (${\rm \delta} V_{\rm i}$). Note that due to small
number statistics the density variations may be non-significant and
thus we will be using $w(r)=1$ whenever $\sigma(w) \ge |1-w(r)|$.

\subsection{BCS sample}
The BCS is the biggest X--ray selected, X--ray
flux--limited compilation covering the extragalactic sky in the
northern hemisphere (${\it \delta}\ge 0^{\circ}$, $|b|\ge
20^{\circ}$). It contains 199 clusters above $S_{\rm lim}=4.45 \times
10^{-12}$erg s$^{-1}$ cm$^{-2}$, in the same energy band as the XBACs,
with $z\le 0.3$ and with X--ray luminosities $\ge 1.25
\times 10^{42}\,h^{-2}\,{\rm erg\,s^{-1}}$.
The BCS list includes not only Abell clusters but also the brightest
Zwicky clusters and others selected on the basis of their X--ray
properties alone. 
It therefore has a significant overlap with XBACs (for ${\it \delta} \ge 
0^{\circ}$) as far as the Abell population is concerned. The above BCS sample
is estimated to be 90 per cent complete (redshift completion is more
than 96 per cent).

\subsection{X--ray cluster selection functions}
Necessary in estimating the local acceleration field from flux--limited
samples is the use of the sample selection function which is determined 
in our case by the cluster X--ray luminosity function,
$\Phi_{\rm x} (L)$.
Ebeling et al. (1997c; 1997b) have recently fitted to the data a
Schechter luminosity function (with parameters given in 
Table 1):
\begin{equation}\label{eq:lf}
\Phi_{\rm x}(L) = A\,\exp\left(-\frac{L}{L_{\ast}}\right)\!L^{-\alpha},
\end{equation}
where $L_{\ast}$ is the characteristic luminosity measured in
$10^{44}\,h^{-2}\, {\rm erg}\;{\rm s}^{-1}$, $A$ being the overall 
normalization of the number--density measured in $h^{3}$ Mpc$^{-3}\,
(10^{44}\,h^{-2}\,{\rm erg}\,{\rm s^{-1}})^{\alpha -1}$, and $\alpha$ is 
the usual power--law index. 

The selection function, defined as the fraction of the cluster number density 
that is observed above the flux limit at some distance $r$,  is: 
\begin{equation}\label{eq:sf}
\phi(r) = \frac{1}{\bar{n}_{\rm c}} \int_{L_{\rm min}(r)}^{L_{\rm
max}} \Phi_{\rm x}(L)\;{\rm d}L.
\end{equation}
with $L_{\rm min}(r)=4\pi r^{2} S_{\rm lim}$ and $L_{\rm max} \simeq
10^{45} h^{-2}\,$erg s$^{-1}$ (due to the form of $\Phi_{\rm x}(L)$ the
above integral is very insensitive to larger values of $L_{\rm max}$). 
The mean number density of the underlying X--ray population of clusters
is found by integrating the luminosity function from the lower to
the upper luminosity limit of the sample:
\begin{equation}\label{eq:den}
\bar{n}_{\rm c}=\int_{L_{{\rm min}}}^{L_{{\rm max}}} \Phi_{\rm x}(L)
\;{\rm d}L.
\end{equation}
Since $L_{\rm min}$, the absolute lower luminosity limit, for the XBACs 
sample is effectively unknown we can estimate it by relating the above 
equation with the observed number density of the optical Abell/ACO sample 
which can be considered as the `parent' population of the XBACs sample.
Using the weighted mean number density of Abell/ACO clusters, corrected for
Galactic absorption ($\bar{n}_{\rm c} \simeq1.85\,(^{+0.6}_{-0.3}) 
\times10^{-5} h^{3}\,{\rm Mpc^{-3}}$)
we obtain $L_{\rm min}=4.3\,(^{+ 3.5}_{- 2.5}) \times 10^{41}\,h^{-2}\,$erg
s$^{-1}$; the uncertainty reflecting the density variations between
the Abell and ACO samples. For the case of the BCS sample, for which 
$L_{\rm min}=1.25 \times 10^{42} \; h^{-2}$ ${\rm erg\;s^{-1}}$, we obtain
that the global mean number density of its parent X--ray cluster population is
$\bar{n}_{\rm c}=5.81\times 10^{-5}\,h^{3}\,$Mpc$^{-3}$, a factor of $\sim$3 
times larger than that of the XBACs sample.

The predicted number of X--ray clusters above $S_{\rm lim}$ and lying within 
a shell between $r$ and $r+\Delta r$, is then:
\begin{equation}\label{eq:Nr}
N(r)=4\pi r^{2} \phi(r) \bar{n}_{\rm c} \Delta r = 4\pi r^{2} \Delta r 
\int_{L_{{\rm min}}(r)}^{L_{{\rm max}}} \Phi_{\rm x}(L) \; {\rm d}L 
\end{equation} 
Note that $N(r)$ is independent of 
$\bar{n}_{\rm c}$ and thus of the uncertainty in $L_{\rm min}$.
Fig. 1 shows the observed number of XBACs clusters as a function of
distance and the predicted one from equation (\ref{eq:Nr}).
The maximum of $N(r)$ turns out to be around $\sim 240h^{-1}\,$Mpc, 
in agreement with the observed distribution. 
If we choose to fit better only the region of 
reliable redshifts ($\mincir 0.1$) we would obtain for the luminosity 
function parameters:
$\alpha \approx 1.25$ and $L_{\ast}\approx 1.2 \times
10^{44}h^{-2}\,$erg s$^ {-1}$ (dashed line in fig. 1), which
although deviate from the nominal values of Table 1, they are within
their 1$\sigma$ uncertainty. The insert of fig. 1 shows the corresponding 
$N(r)$ of the optical Abell/ACO cluster, corrected for Galactic absorption 
(stars) and the theoretical curve for the range of Abell and ACO densities.
 
Fig. 2 presents the observed BCS $N(r)$ distribution, for 
$|b|\ge 30^{\circ}$, and the corresponding theoretical one (equation
\ref{eq:Nr}).
The maximum of the selection function turns out to be at $\sim
140h^{-1}\,$Mpc, followed by a long tail towards larger depths. This
early maximum ensures that the BCS function is dominated by relatively 
local clusters, more so than the corresponding XBACs sample. 

Note that we will limit our dipole analysis within 240$h^{-1}\,$Mpc
to avoid possible systematic effects due to the low number of the 
observed X--ray clusters and due to uncertainties in the $m_{10}-z$
based cluster redshifts that dominate above this depth.

\section{Cluster Dipole}
We will not present all the details of the method used to calculate the
peculiar gravitational acceleration induced by some mass tracer on the 
observer since such can be found in many recent articles 
(cf. Tini-Brunozzi et al. 1995; Kolokotronis et al. 1996 and
references therein).
Briefly, we employ the method of moments to quantify the distribution of
clusters around the LG and we correct them 
for the effects of galactic absorption using a spherical harmonic expansion of
the cluster surface number density and a combined mask to take into account 
the depletion of clusters for $|b|<13^{\circ}$ and a cosec$b$ absortion law
above this latitude limit (see Plionis \& Valdarnini 1991 and the
appendix of Tini-Brunozzi et al. 1995 for details).
We then estimate the gravitational acceleration induced on the LG from the 
distribution of X--ray clusters (cf. Miyaji \& Boldt 1990; Plionis et al. 
1993) by:
\begin{equation}\label{eq:vel}
{\bf V}_{\rm g}(r) = H_{\circ} r\,\frac{\bf D}{{\cal M}}(\le r) \;\;
\mbox{ for $r\ge R_{\rm conv}$}
\end{equation}
where ${\bf D} = \sum {\cal W}_{\rm i} r_{\rm i}^{-2} \hat{{\bf r}}_{\rm i}$
is the dipole, ${\cal M} = \sum {\cal W}_{\rm i} r_{\rm i}^{-2}$ is the
monopole, ${\cal W}_{\rm i} (\propto w_{\rm i}\, \phi_{\rm i}^{-1} 
M_{\rm i})$ are the cluster weights with $M_{\rm i}$ an estimate of
the cluster mass, $\phi_i$ the cluster selection function and $w_i$ the 
Abell/ACO relative weight (see equation \ref{eq:weight}). 
$R_{\rm conv}$ is the dipole convergence depth, the depth beyond which the 
distant density inhomogeneities do not affect the dynamics of the observer and
should therefore be within the effective depth of the 
catalogue in order to obtain the correct estimate of the local acceleration 
field. 
Using the definition of the monopole (${\cal M} = \int \rho(r) r^{-2}
{\rm d}V$) and linear perturbation theory (cf. Peebles 1980) we can recover 
from equation (\ref{eq:vel}) the more familiar form: 
\be\label{eq:ulg}
{\bf u}_{\rm LG}(r) = \beta_{\rm c_{x}} {\bf D}(r)/4 \pi \bar{n}_{\rm c} 
= \beta_{\rm c_{x}}\,{\bf V}_{\rm g}(r)\;.
\ee
where $\beta_{\rm c_{x}} \equiv \Omega_{\circ}^{0.6}/b_{\rm c_{x}}$ and 
$b_{\rm c_{x}}$ is the X--ray cluster to underlying mass bias factor. 
Note that we will be using two mass weighting schemes; one in
which we will assume each cluster to contribute equally ($M=1$) and one 
in which the mass is proportional to the X--ray
luminosity ($M \propto L_{\rm x}^{5/11}$). This relation results from
the assumption of hydrostatic equilibrium, $T \propto M^{2/3}$, and from
{\small GINGA} observations which indicate that $L_{\rm x} \propto T^{3.3}$
(cf. Arnaud 1994 and references therein).   

The sparseness, however, with which the flux--limited sample of clusters trace 
their underlying parent cluster population introduces shot-noise 
(discreteness effects) in their dipole estimates which increase with distance.
Kolokotronis et al. (1997) found that the 
enhancement of the underlying true cluster X--ray dipole due to shot-noise 
and the loss of dipole signal due to the flux--limited nature of the sample 
work in opposite directions, tending to counteract each other. Therefore, 
although we estimate the magnitude of the 1D shot-noise dipole, using 
the formalism developed in Strauss et al. (1992), i.e. 
$|{\bf D}|_{\rm sn, 1D}^{2} \approx 1/3 \sum
\phi^{-1}_{\rm i} r^{-4}_{\rm i} (\phi^{-1}_{\rm i} - 1)$, finding it to be 
about $\sim 30$ per cent of the dipole signal, 
we do not attempt to correct the {\em raw} XBACs dipole for such effects 
(see also Hudson 1993 and Kolokotronis et al. 1997 for alternative 
shot-noise definitions).

\subsection{Dipole Results}
In fig. 3 we present the cluster dipole (based on both mass weighting 
schemes) for the XBACs sample (triangles) as well as for the optical 
Abell/ACO sample (squares). 
It is evident that both samples excibit a very similar dipole
profile with significant contributions from depths $\gg 100h^{-1}\,$Mpc, which
validates the previous results based only on the optical sample (Plionis \&
Valdarnini 1991; Scaramella, Vettolanni \& Zamorani 1991; Branchini \& 
Plionis 1996). However, the XBACs dipole is systematically lower, by $\sim 
20$ per cent (for the equal mass weighting case), than the optical cluster 
dipole. Although this could be intrinsic, implying that 
the optical dipole is artificially enhanced by projection effects (cf. 
Sutherland 1988; Peacock \& West 1992), such an explanation is not 
corroborated by the correlation function analysis of the XBACs sample which 
provides a large correlation length, roughly consistent with that of the 
optical Abell/ACO sample (Edge, private communication). 
An alternative explanation of the lower XBACs dipole 
amplitude with respect to the optical one, is a possibly artificial exclusion
from the XBACs catalogue of nearby clusters ($\mincir 50 - 60h^{-1}\,$Mpc) 
which naturally play a key role in shaping the local acceleration field.
In fact, from the 8 Abell/ACO clusters within 60$h^{-1}\,$Mpc not included
in the XBACs sample, three (A3565, A3574 and A347), although detected in 
{\small RASS}, have been excluded because 
of suspision that their X--ray emission is mostly of non-cluster origin.
If we include by hand these three clusters, then the XBACs ($M=1$) dipole 
increases substantially, reducing the difference with the
optical dipole from $\sim$ 20 to $\sim$ 10 per cent.
This reduced discrepancy could be further gapped if we take into account 
the results of Kolokotronis et al. (1997) who found, using numerical 
experiments, that the cluster X--ray flux--limited and unity weighted ($M=1$)
dipole will underestimate by $\sim$ 15 per cent the underline cluster
dipole if it has significant contributions from large depths  
($\gg 100h^{-1}\,$Mpc). 

However, using the more natural luminosity weighting scheme ($M \propto 
L_{\rm x}^{5/11}$) we find an even lower amplitude of the
XBACs dipole with respect to the unity weighted one, although their
dipole profiles are very similar. The amplitude of the luminosity
weighted X--ray cluster dipole is by $\sim$ 35 per cent less than its
optical counterpart and although the $\sim 25$ per cent gap could be
bridged, as discussed above, it seems that the X--ray cluster dipole
is intrinsically less, by $\magcir 10$ per cent, than the optical
cluster dipole. This difference, if it is indeed intrinsic,
corresponds to an optical to X-ray cluster bias factor $b_{\rm o,x}(\equiv
b_{\rm o}/b_{\rm x}) >1$. 

In order to further investigate these points we will attempt to fill 
in the lack of local information by using the BCS clusters, which better 
trace the local volume (see fig. 2) and compare the XBACs and BCS dipoles in
their common region ($\delta > 0^{\circ}$, $|b|>20^{\circ}$). 
One's hope is that at the convergence depth of the XBACs dipole the relative 
fluctuation between the BCS and XBACs dipoles will 
reflect those of the whole-sky dipole, and therefore we could infer a better 
estimate of the final X--ray cluster dipole amplitude. 

\subsection{Comparing the XBACs and BCS dipoles}
The northern XBACs sample (hereafter XBACs-n) contains 113 clusters
out of which 112 belong to the BCS sample as well, with A2637 being
the sole exception (for details see section 10 of Ebeling et
al. 1997a).  
Furthermore, although 65 per cent of the clusters of the two  
samples are common, it is not straight-forward that they should trace
similarly the northern hemisphere dipole since (a) the BCS is governed
by a significantly different $\Phi_{\rm x}(L)$ which results in a  different
$N(r)$ distribution (see figs. 1 and 2) and 
(b) only $\sim 33$ and $\sim 53$ per cent of the clusters are common
within the interesting regions ($\sim 100$ and $\sim 200 \;h^{-1}$ Mpc 
resepectively). 

For this comparative work we will correct the
raw dipole estimates for shot-noise errors since the 
two samples should trace the same underlying distribution but with different
densities and selection functions. 
We plot in the lower panel of fig. 4 the fluctuations of the unity 
weighted XBACs-n and BCS dipoles ($\delta V/V \equiv V_{\rm XBACs-n}-
V_{\rm BCS}/V_{\rm BCS}$) 
including (dashed line) and excluding Virgo (solid line). 
We also plot (upper panel) the misalignment angle between the XBACs-n and 
BCS dipoles at each distance bin for the the luminosity weighted one 
(solid line) and for the unity-weighted dipole case (long dashed line) 
excluding Virgo. 
The short dashed line corresponds to the luminosity weighted dipoles
but including in the BCS sample the Virgo cluster. 

The most significant results of this analysis are:
\begin{itemize}
\item Comparing consistently the BCS and XBACs-n dipoles, i.e. excluding Virgo
from the BCS sample since by construction it is absent from the XBACs, we 
find that both X--ray samples have very similar dipole
shapes, with small amplitude differences ($\delta V/V \mincir 0.05$) and 
$\delta\theta \mincir 14^{\circ}$ at scales $\magcir R_{\rm conv}$.
Note that the $\delta\theta$ values are uncorrected for the misalignment
induced by the shot noise dipole, which is roughly $\sim 10^{\circ}$.
\item If one takes into account the above value of the relative velocity 
difference between the BCS and the XBACs-n one would further reduce the 
apparent gap between the XBACs all-sky and the optical Abell/ACO dipole 
but by not more than $\sim 5$ per cent.
\item Similarly with the XBACs also in the BCS case the luminosity weighted 
dipole is by $\sim 20$ per cent lower than the corresponding unity weighted 
one.
\item Including Virgo in the BCS sample we find, as expected, that it plays 
a significant role in shaping the X--ray dipole, with relative contribution of 
$\sim 10 - 15$ per cent which corresponds to an average Virgocentric infall
velocity of $\sim 160 \pm 40$ km/sec (were we have 
weighted twice the luminosity based results).
\end{itemize}

\section{Estimating the density parameter $\beta$}
The good alignment (within $\sim 25^{\circ}$) between the XBACs and CMB dipoles
indicate that the XBACs clusters trace the large-scale mass density 
field and that they can therefore be used to estimate the cosmological 
$\beta$ parameter by relating the X--ray cluster dipole to the 
Local Group peculiar velocity (equation \ref{eq:ulg}).
However since the Virgo cluster is not included in the Abell sample,
due to its proximity and thus low surface density, we must exclude from
the LG peculiar motion the Virgocentric Infall. Equation (\ref{eq:ulg})
then becomes:
\be \label{eq:inf} 
{\bf u}^{'}_{\rm LG} =  {\bf u}_{\rm LG} - {\bf V}_{\rm inf} =
\beta_{\rm c_{x}} \; |{\bf V}_{\rm g}| \;.
\ee
Using $V_{\rm inf} \simeq 170$ ${\rm km\;s^{-1}}$ 
we find $|{\bf u}^{'}_{\rm LG}|\simeq 500$ ${\rm km\;s^{-1}}$, pointing 
towards $(l,b)= (265^{\circ},15^{\circ})$.

\noindent
The cluster redshift is related to its the comoving distance by:
\be
c z = H_{\circ} r + [{\bf v}_{\rm p}(r) - {\bf u}_{\rm LG}] \cdot
\hat{\bf r}
\ee
and since the last term of this equation is $\neq 0$,  
redshift-space distortions will tend to enhance the dipole amplitude 
(Kaiser 1987). In an attempt
to derive the optical cluster dipole free of such distortions,
Branchini \& Plionis (1996) used a density reconstruction algorithm to
predict the real-space positions of the optical
Abell/ACO clusters. They found that redshift space distortions
($r.s.d$ hereafter)
enhance the real-space optical cluster dipole by $\sim$ 23 per cent.  
In order to correct the XBACs dipole for such effects we attempt
to minimize $r.s.d$ using a simple model of the peculiar velocity field. 
Since we observe in the local Universe a coherent bulk
flow of high amplitude (cf. Dekel 1994, 1997; Strauss \& Willick 1995), we
split the cluster peculiar velocities in a component of a bulk flow
and a local non-linear component as follows:
\begin{equation}\label{eq:pecs}
{\bf v}_{\rm p}(r) = {\bf V}_{\rm bulk}(r) + {\bf V}_{\rm nl}(r)
\end{equation}
Applying equation (\ref{eq:pecs}) to the Local Group and using
equation (\ref{eq:inf}) we have that ${\bf V}_{\rm bulk}(0) = {\bf
u}^{'}_{\rm LG}$ from which it is evident that locally the bulk flow
component dominates over that of the infall.
This fact may be reversed for galaxies at large distances but in any case 
at such distances we have $({\bf v}_{\rm p} \cdot \hat{\bf r})/ cz \ll 1$ 
and thus {\em r.s.d} should not significantly affect the dipole. 
We therefore use the approximation ${\bf v}_{\rm p}(r) = {\bf V}_{\rm
bulk}(r)$, where the bulk flow profile, as a function of distance, 
is given by Dekel (1994, 1997) and by Branchini, Plionis \& Sciama (1996),
and its direction is taken to be that of ${\bf V}_{\rm bulk}(0)$.
Note, however, that there have been measurments of the bulk velocity with very
different results from the above in direction as well as in amplitude 
(Lauer \& Postman 1994). 
The reality, however, of these results have been questioned 
by different studies (cf. Giovanelli et al 1996; Hudson \& Ebeling 1996).

Our results are completely compatible with those of the full
reconstruction of Branchini \& Plionis (1996); the redshift-space
XBACs dipole is enhanced by $\sim$ 20 per cent with respect to the
corrected ({\em real-space}) dipole. The main dipole results and the 
corresponding values of $\beta_{\rm c_{x}}$ (using equation \ref{eq:ulg}) are 
presented in Table 2. 
Taking into account a possibile $\sim$ 20 per cent artificial
decrease of the X-ray dipole (see discussion in 3.1) and averaging over the
different determinations (weighting twice the more physical luminosity
weighted results) we obtain 
$$\beta_{\rm c_{x}} \simeq 0.24 \pm 0.05 \;.$$
Note that from the optical Abell/ACO cluster dipole Branchini \& Plionis
(1996) found $\beta_{\rm c_{o}} \approx 0.21$. Furthermore, Branchini et al. 
(1997) comparing the real-space optical cluster density field 
(within $\sim 70h^{-1}\,$Mpc) with the corresponding POTENT-Mark III 
field found $\beta_{\rm c_{o}} \approx 0.20 \pm 0.06$.
The difference between their $\beta_{\rm c}$ value and the present analysis
could be attributed to an optical to X--ray cluster bias factor $b_{\rm o,x}
\simeq 1.2$.

\section{Conclusions}

We have estimated the X--ray cluster dipole, using the whole-sky XBACs sample
and the BCS sample which covers the northern hemisphere. We have found that:
\begin{itemize} 
\item[(a)] The relative contributions to the LG 
acceleration field, from different depths, is readily provided by the XBACs 
dipole analysis and supports the conclusions drawn from the optical 
Abell/ACO cluster analysis of significant dipole contributions
($\sim 30 - 40$ per cent of total) from scales $\sim 130 - 160h^{-1}\,$Mpc. 
Furthermore, the XBACs and BCS clusters trace equally well the same 
dipole structure and thus the large-scale density field.
\item[(b)] Using a model of the large-scale peculiar velocity field and 
correcting 
for redshift space distortions, we find that the {\em real-space} X--ray 
cluster dipole is reduced by $\sim$ 20 per cent, a value consistent with 
the outcome of a full reconstruction of the optical cluster density field.
\item[(c)] Although the `zero-point' of the X--ray cluster dipole cannot be 
unambigiously determined from the present analysis we find that the true 
underlying X--ray cluster dipole is intrinsically 
lower than the corresponding optical cluster dipole by  $\sim 10$
to 30 per cent, depending on whether the X--ray clusters are weighted equally 
or $\propto L_{\rm x}^{5/11}$ and on whether one assumes that the
observed X--ray emission of a few nearby clusters (A3565, A3574 and
A347) is of non-cluster origin. 
\item[(d)] Relating the X-ray cluster dipole with the LG peculiar velocity
we find $\Omega_{\circ}^{0.6}/b_{\rm c_{x}} \simeq 0.24\pm 0.05$, which 
combined with recent determinations based on comparing the optical cluster
density and velocity fields with the corresponding POTENT-Mark III
fields, imply a relative optical to X--ray cluster bias factor of $b_{\rm o,x}
\simeq 1.2$.
\item[(e)] The Virgo cluster contributes about $\sim 10 - 15$ per cent of the 
overall X-ray cluster dipole which corresponds to an average Virgocentric 
infall velocity of $\sim 160 \pm 40$ km/sec.
\end{itemize}

\vspace {1cm}

\acknowledgments
We greatly thank Harald Ebeling for many useful discussions, for
kindly giving us the XBACs \& BCS luminosity functions and the BCS
sample prior to publication and for always answering meticulously our
many questions. We also acknowledge fruitful discussions with Alastair
Edge on the statistical properties of the XBACs and BCS samples.

\newpage

\section*{Tables}

\begin{table}[h]
\centering
\caption[] {Parameters for the XBACs and BCS X--ray luminosity function, 
using $H_{\circ}=100 \; h$ km s$^{-1}$ Mpc$^{-1}$.}
\tabcolsep 40pt
\begin{tabular}{ccc} \\ \hline
Parameters &  XBACs  &  BCS \\ \hline \\
$A$   &  $1.955\times10^{-6}$  &$1.246\times10^{-6}$ \\  
$\alpha$   &  $1.21^{+0.12}_{-0.13}$   &  $1.85 \pm 0.09$   \\
$L_{\ast}$ &  $1.048^{+0.17}_{-0.14}$ & $2.275^{+0.515}_{-0.373}$ \\ \hline
\end{tabular}
\end{table}

\begin{table}[h]
\centering
\caption[] {XBACs Dipole Parameters and the corresponding values of
$\beta_{\rm c_{\rm x}}$ at $r=200 \; h^{-1}\,$Mpc. Note that ${\rm
\delta}\theta_{\rm cmb}$ is the dipole misalignment angle from the
CMB dipole direction corrected for a 170 km$\;{\rm s}^{-1}$
Virgocentric infall.}
\tabcolsep 17pt
\begin{tabular}{ccccccc} \\ \hline
frame &$M$ &  $|{\bf V}_{\rm g}|$ (km/sec) & $l^{\circ}$ &
$b^{\circ}$ & ${\rm \delta}\theta_{\rm cmb}$ & $\beta_{\rm c_{x}}$ \\ \hline \\
$z$-space & 1 & 2710 & 269 & 0  & 17$^{\circ}$  & 0.19 \\
$z$-space &$L_{\rm x}^{5/11}$ & 2100   & 275 & 15 & 19$^{\circ}$ & 0.24 \\ \\
{\em real}-space & 1 & 2250 &  255 & -7 & 25$^{\circ}$ & {\bf 0.22} \\
{\em real}-space &$L_{\rm x}^{5/11}$ & 1750 & 251 & 10 & 15$^{\circ}$ & 
{\bf 0.29} \\ \hline
\end{tabular}
\end{table}

\newpage

\begin{figure}
\mbox{\epsfxsize=12 cm\epsffile{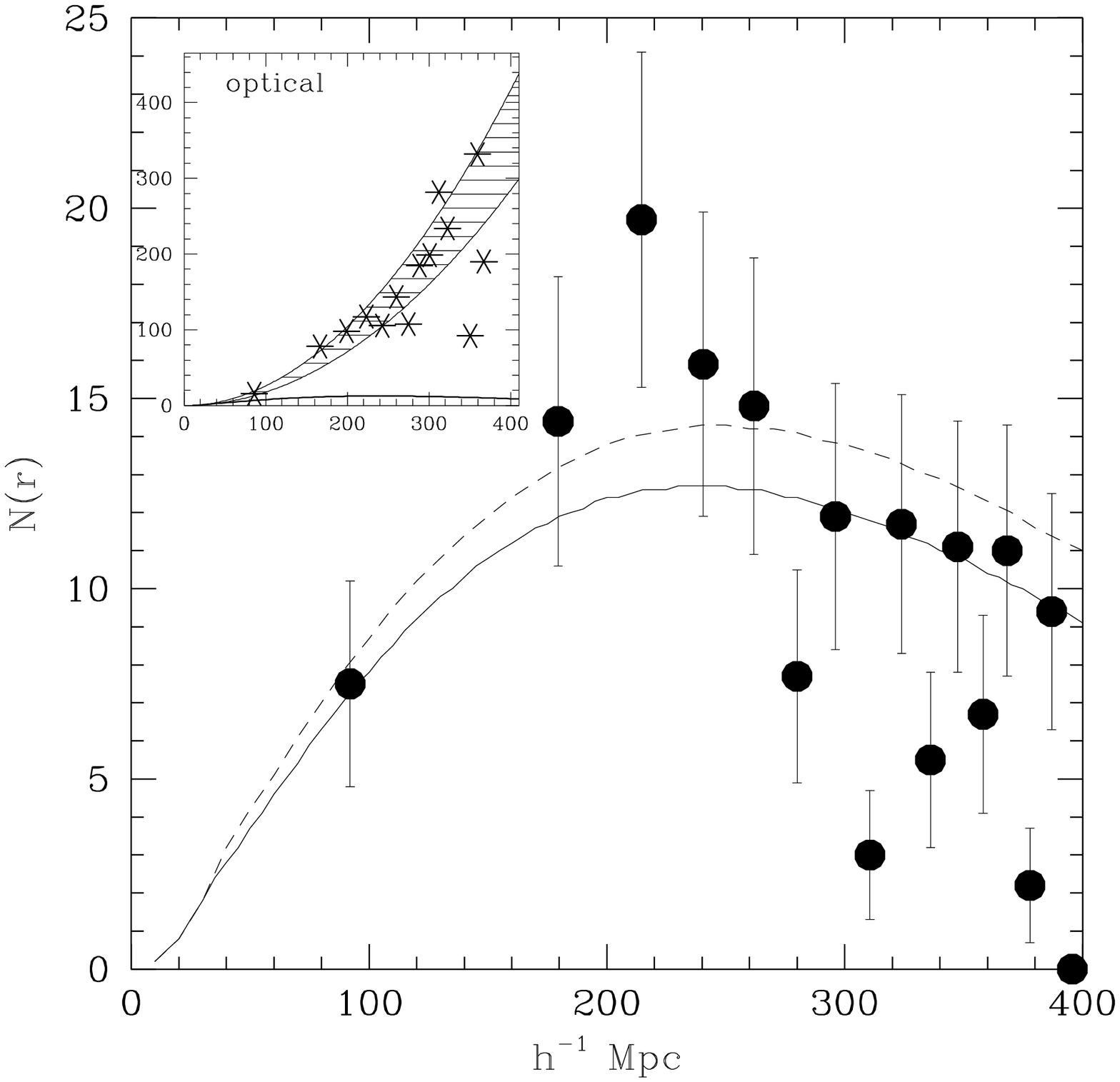}}
\caption{Observed $N(r)$ distribution of XBACs clusters, corrected for
Galactic absorption, and its Poisson uncertainties.
The predicted distribution via equation (\ref{eq:Nr}) is shown as a
solid line. 
The insert shows the observed $N(r)$ distribution of the optical Abell/ACO 
clusters (stars), with the shaded area corresponding to the homogeneous 
case (i.e. $\phi=1$) for densities, $\bar{n}_{\rm c}$, between the Abell and 
ACO values.} 
\end{figure}

\begin{figure}
\mbox{\epsfxsize=12 cm\epsffile{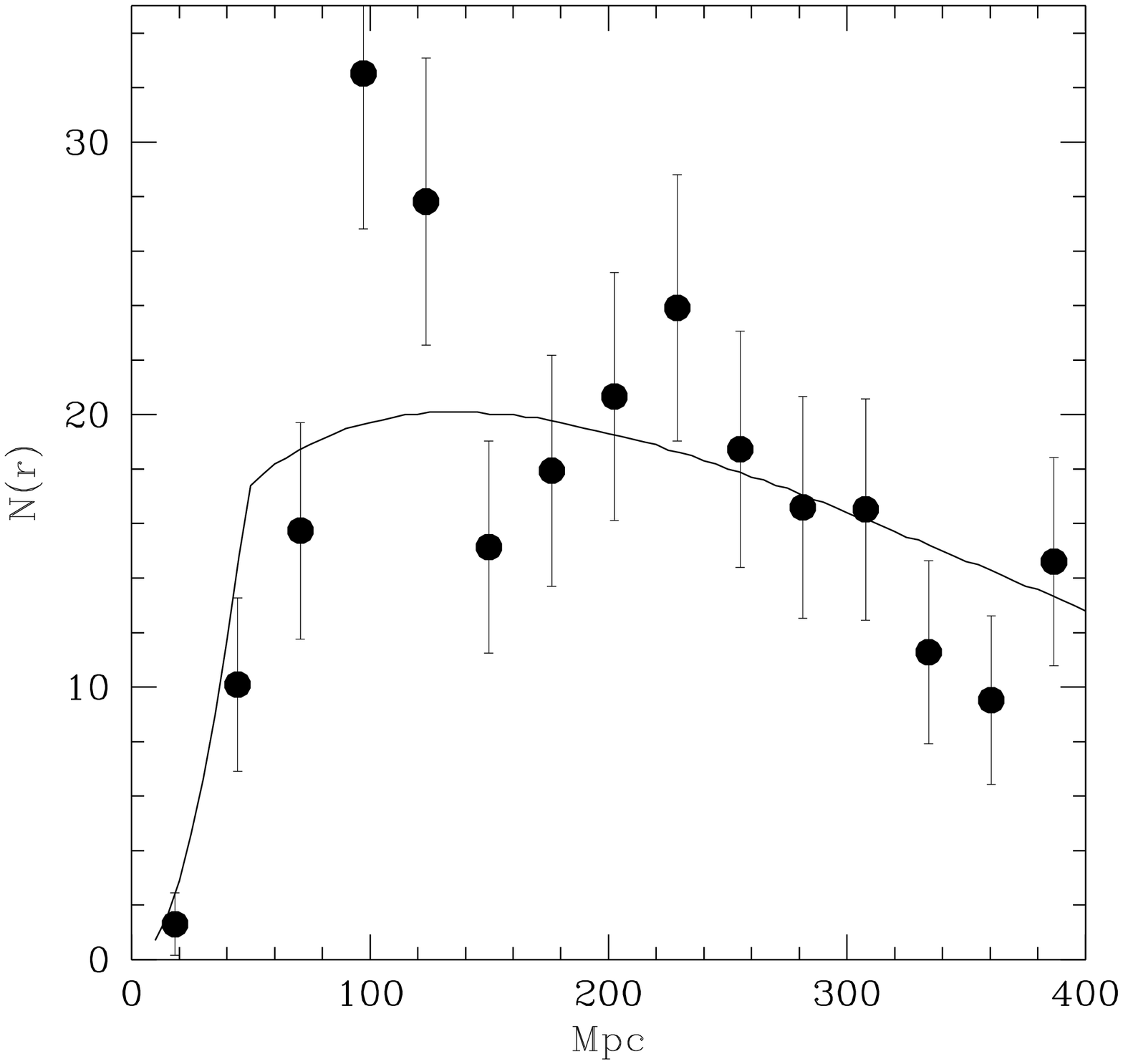}}
\caption{Observed $N(r)$ distribution of BCS clusters, corrected for
Galactic absorption, and its Poisson uncertainties (points). The predicted 
distribution via equation (\ref{eq:Nr}) is shown as a solid line.}
\end{figure}

\begin{figure}
\mbox{\epsfxsize=12 cm\epsffile{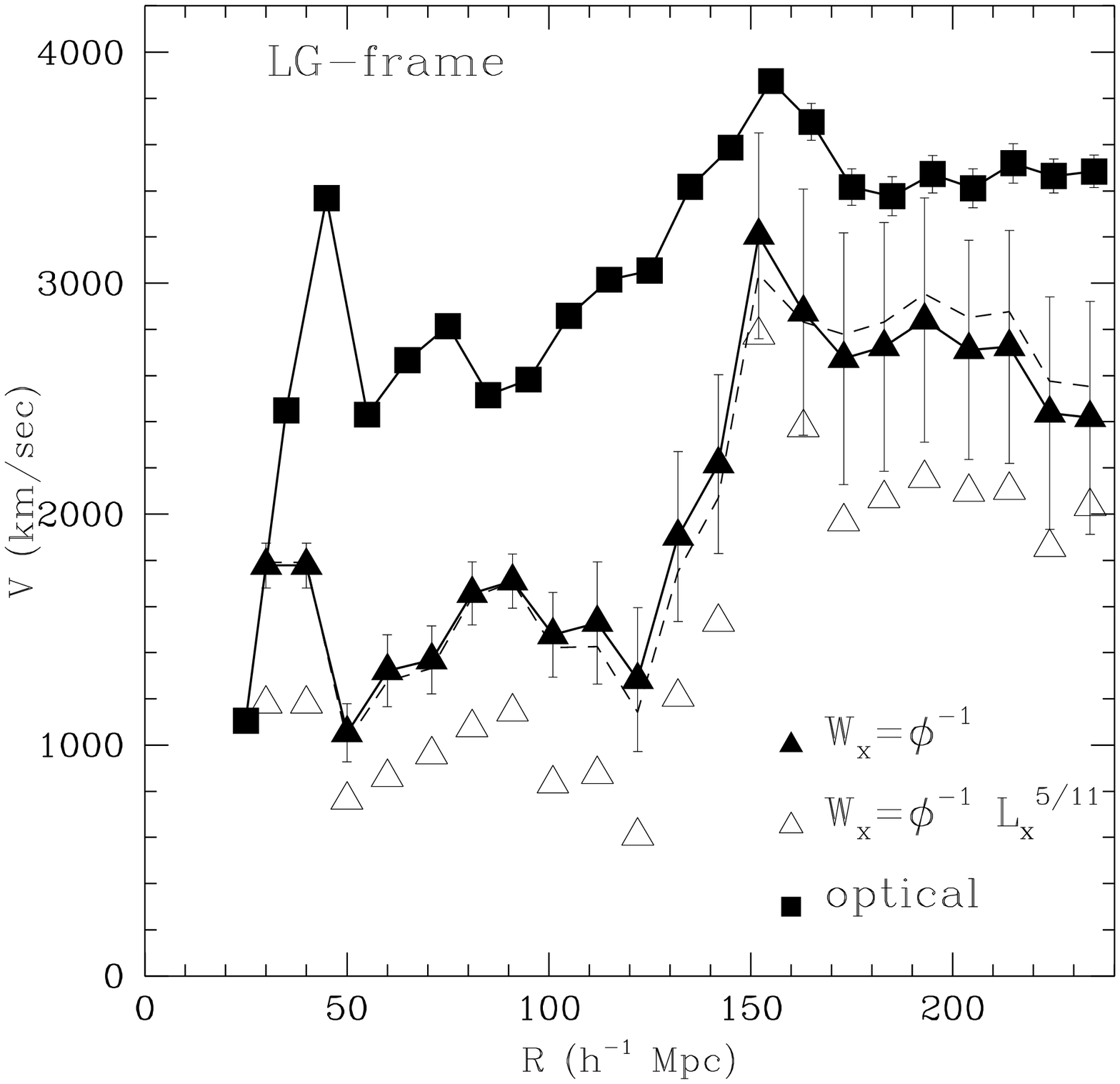}}
\caption{Abell/ACO optical (squares) and XBACs (triangless) dipole. The 
weighting scheme used is indicated. 
Errorbars are 1$\sigma$ uncertainties due to 
different bin sizes used to homogenize the Abell and ACO number densities 
(see equation 2), while the dashed line is the case with no relative weighting
between the Abell and ACO portions of XBACs.}
\end{figure}

\begin{figure}
\mbox{\epsfxsize=12 cm\epsffile{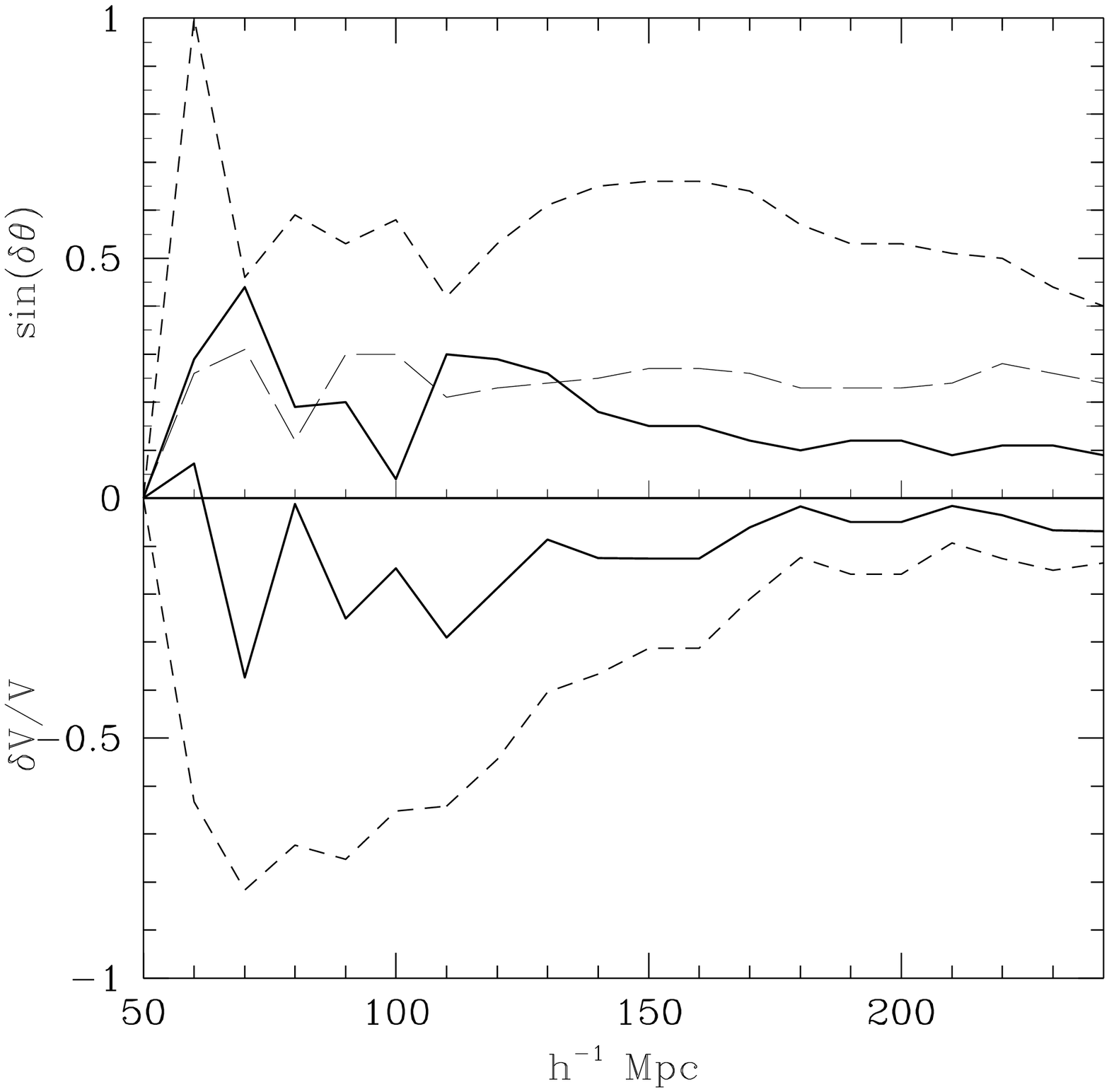}}
\caption{Dipole amplitude variation ($\delta V/V$) between BCS and XBAC-n
for the unity weighting scheme excluding (solid line) and including Virgo 
(dashed line).
Misalignment angles, $\delta\theta$, where the solid  and short-dashed lines 
correspond to the $L_{\rm x}$ weighted case excluding and including Virgo, 
respectively. The long dashed line corresponds to the unity weighted case, 
excluding Virgo.}
\end{figure}

\end{document}